\documentclass[aps,prl,superscriptaddress]{revtex4}
\usepackage{graphicx}
\usepackage{textcomp}
\usepackage{mathptmx}
\begin{document}

\title{Vortex conveyor belt for matter-wave coherent splitting and interferometry}

\author{Jixun Liu}
\email{liujixun@buaa.edu.cn}
\affiliation{Institute of Optics and Electronics Technology, School of Instrumentation Science and Opto-electronics Engineering, Beihang University, Beijing, 100191, China}
\affiliation{Midlands Ultracold Atom Research Centre, School of Physics and Astronomy, University of Birmingham, Edgbaston, Birmingham, B15 2TT, United Kingdom}
\author{Xi Wang}
\affiliation{Midlands Ultracold Atom Research Centre, School of Physics and Astronomy, University of Birmingham, Edgbaston, Birmingham, B15 2TT, United Kingdom}
\author{Jorge Mellado Mu\~noz}
\affiliation{Midlands Ultracold Atom Research Centre, School of Physics and Astronomy, University of Birmingham, Edgbaston, Birmingham, B15 2TT, United Kingdom}
\author{Anna Kowalczyk}
\affiliation{Midlands Ultracold Atom Research Centre, School of Physics and Astronomy, University of Birmingham, Edgbaston, Birmingham, B15 2TT, United Kingdom}
\author{Giovanni Barontini}
\affiliation{Midlands Ultracold Atom Research Centre, School of Physics and Astronomy, University of Birmingham, Edgbaston, Birmingham, B15 2TT, United Kingdom}

\begin{abstract}
We numerically study a matter wave interferometer realized by splitting a trapped Bose-Einstein condensate with phase imprinting. We show that a simple step-like imprinting pattern rapidly decays into a string of vortices that can generate opposite velocities on the two halves of the condensate. We first study in detail the splitting and launching effect of these vortex structures, whose functioning resembles the one of a conveyor belt, and we show that the initial exit velocity along the vortex conveyor belt can be controlled continuously by adjusting the vortex distance. We finally characterize the complete interferometric sequence, demonstrating how the phase of the resulting interference fringe can be used to measure an external acceleration. The proposed scheme has the potential to be developed into compact and high precision accelerometers.
\end{abstract}

\maketitle

\section*{Introduction}

Atom interferometry is one of the pillars of the emerging quantum technologies with applications ranging from the measurement of fundamental physical constants\cite{parker2018measurement, thomas2017determination} to tests of the equivalence principle\cite{dimopoulos2007testing, rosi2017quantum} and the detection of gravitational waves\cite{dimopoulos2008atomic, kovachy2015quantum, canuel2018exploring}. Additionally, atom interferometers are promising candidates for navigation and geophysics surveys as they have shown excellent performance in the field of acceleration and rotation measurements\cite{menoret2017transportable, menoret2018gravity}. Atom interferometers can be implemented with collimated atomic beams\cite{giltner1995atom, keith1991interferometer} or cold atoms\cite{kasevich1991atomic}, for which coherent splitting and recombination of the matter wave can be achieved by using  microfabricated gratings\cite{keith1991interferometer}, Bragg scattering\cite{giltner1995atom} or Kapitza-Dirac scattering\cite{giltner1995atom, rasel1995atom} from standing light waves, or stimulated Raman transitions\cite{kasevich1991atomic}. Nowadays free-falling atom interferometers can reach record accuracy up to of 3.9~$\mu$Gal \cite{freier2016mobile}, 
surpassing other state-of-the-art techniques. However, as they are reaching their ultimate performances, it's quite challenging to make any further improvement. 
For example, to improve their sensitivity, it is necessary to increase the interrogation time, but this comes at the price of increasing sizes of the atomic clouds or of the atomic beams, making it harder to manipulate the atoms and extract the useful signal. More recently, advanced cooling techniques have allowed to achieve interrogation time of 2.3~s \cite{dickerson2013multiaxis}, however as the atoms are in free fall, the vacuum enclosure has to be on the order of 10~m length\cite{dickerson2013multiaxis}. In many applications, like inertial navigation and geodesy, such large scale apparatuses are not practically feasible. In addition, vibrations\cite{hu2013demonstration}, wave-front distortion\cite{gauguet2009characterization}, and detection noise \cite{hu2013demonstration} are other known limiting factors in the performances of these atom interferometers.

In the constant quest for more sensitive and portable devices, a new generation of interferometers based on trapped and guided atomic samples is emerging. As the atoms are held against gravity by the confining potentials, a trapped and guided interferometer promises to provide very long interrogation time within a small apparatus size. This is helpful to suppress phase shifts due to stray fields and mechanical vibrations\cite{wu2007demonstration} and increase the sensitivity. Besides, it allows to have precise control over the atomic wavefunction, which makes trapped interferometers ideal sensors for inertial navigation \cite{schumm2005matter, kasevich2002coherence, kreutzmann2004coherence}, where a large dynamical range is desirable, and for testing (sub-)gravitational and surface-generated forces \cite{dimopoulos2003probing, alauze2018trapped}, due to the high spatial resolution. The many advantages offered by trapped interferometers bear the promise to deliver a new generation of devices able to exceed the capabilities of the free-space counterparts \cite{burke2008confinement}. Additionally, they allow the use of Bose-Einstein condensates (BECs), that are particularly appealing systems for the realization of precise interferometers due to their properties of macroscopic phase coherence \cite{andrews1997observation}. For example, besides Bragg scattering \cite{kovachy2015quantum, wang2005atom}, and Kapitza-Dirac scattering \cite{sapiro2009atom}, the BECs can also be coherently split by deforming a single-well potential to a double-well potential, which can be realized by changing the optical trap \cite{shin2004atom}, or by applying radio-frequency \cite{schumm2005matter} or microwave fields \cite{guarrera}. Additionally, the possibility of controlling the interactions, that are often viewed as a major drawback, allows to engineer entangled states \cite{riedel2012atom} or to use bright solitons \cite{marchant2013controlled, mcdonald2014bright}, further enhancing the sensitivity of the device.

Phase imprinting methods have been proposed to manipulate the phase of the BEC with optical potentials \cite{gajda1999optical}. The manipulation can be almost arbitrary in two dimensions using the widely employed spatial light modulators  \cite{becker2008oscillations, denschlag2000generating}. This powerful technique has been mainly used to study vortices \cite{gajda1999optical} and solitons \cite{becker2008oscillations, denschlag2000generating, meyer2017observation} in BECs but has also potential for applications in atom interferometry \cite{ruschhaupt2009momentum}. For example, it has been shown that interference can be produced in momentum space by imprinting part of a trapped quasi-one-dimensional BEC with a detuned laser \cite{ruschhaupt2009momentum}, similarly to the generation of solitons \cite{becker2008oscillations, denschlag2000generating}. In this work, we study an alternative and simple way to build a trapped matter-wave interferometer using phase imprinting. In our scheme, two opposite velocities are imprinted on two halves of a trapped BEC. We find that this imprinting generates a string of vortices along the central line of the BEC. Our numerical simulations show that the vortices effectively act as a conveyor belt which moves the atoms on its two sides in opposite directions. As a result, the BEC is coherently split into two clouds that, after half oscillation in the trapping potential, recombine in the centre, forming an interference pattern. We demonstrate that the phase of the interference fringe can be used to infer the magnitude of an external acceleration over a large dynamical range.

\begin{figure}[!t]
\centering
	\includegraphics[width=0.45\textwidth]{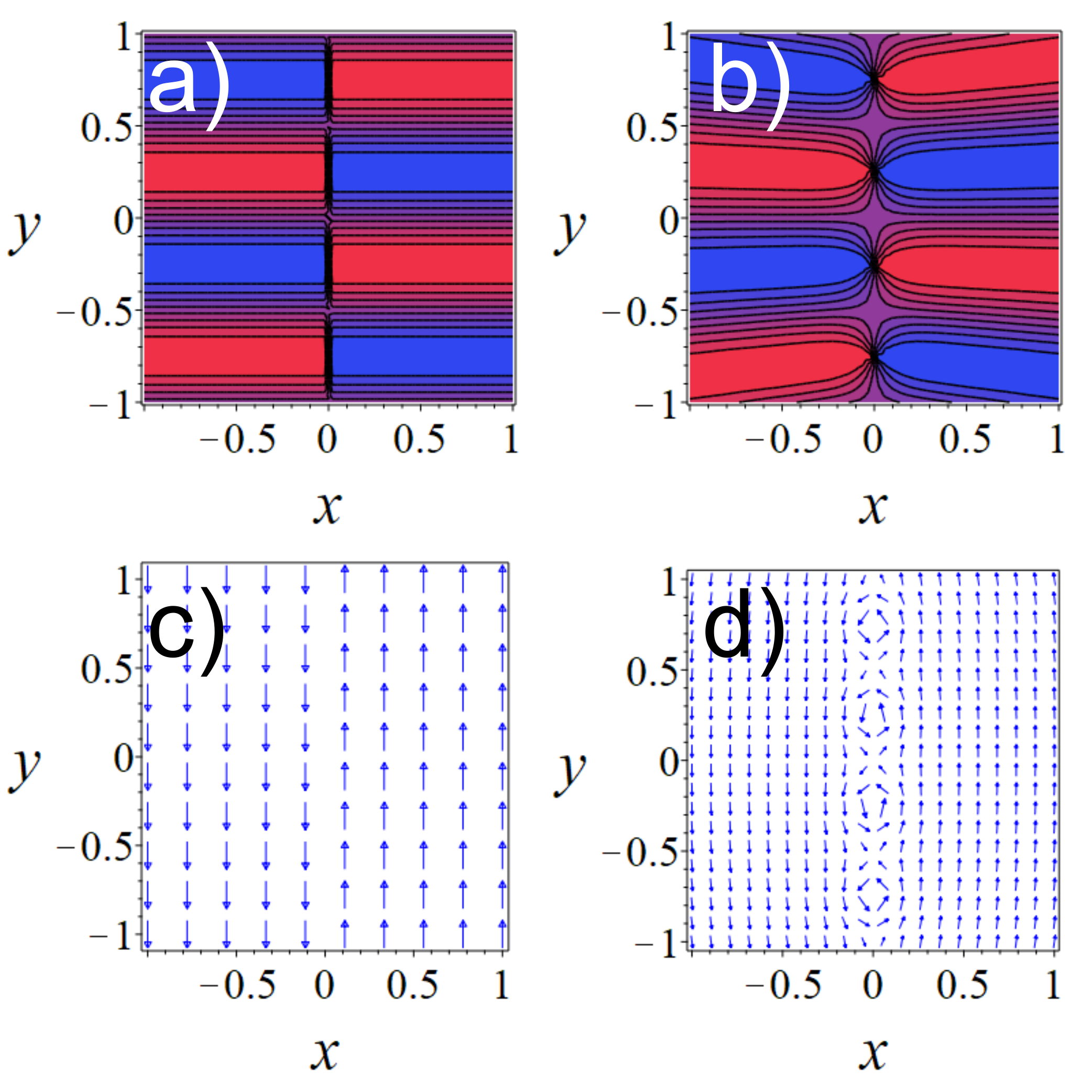}
	\caption{a) Phase pattern resulting from the imprinting of Eq.~(\ref{eq:imprintvelocity}) with $l=1$. b) Phase pattern generated by  Eq.~(\ref{eq:imprintedphase}) with $d=0.25$. c) and d) are the corresponding velocity fields. }
\label{fig1}
\end{figure}

\section*{Results}

Superfluidity is one of the most spectacular consequences of Bose-Einstein condensation \cite{dalfovo1999theory}. At the mean field level, the trapped BECs can be well described by the Gross-Pitaevskii equation, which is a nonlinear Schr\"odinger equation:
\begin{equation}
i\hbar\frac{d}{dt}\Phi(\mathbf{r},t) = \Big(-\frac{\hbar^2\nabla^2}{2m} +V(\mathbf{r})
+\xi|\Phi(\mathbf{r},t)|^2\Big)\Phi(\mathbf{r},t), \label{eq:GPE}
\end{equation}
where $\Phi(\mathbf{r},t)$ is the wave function of the condensate (order parameter), $m$ is the mass of the atom, $V(\mathbf{r})$ is the trapping potential, and $\xi = 4 \pi \hbar^2 a/m$ in which $a$ is the \textit{s}-wave scattering length. The complex order parameter $\Phi(\mathbf{r},t)$ can be written as
\begin{equation}
\Phi(\mathbf{r},t) = \sqrt{n(\mathbf{r},t)} e^{iS(\mathbf{r},t)}, \label{eq:modulus and phase}
\end{equation}
where $n(\mathbf{r},t)$ is the condensate density distribution, and $S(\mathbf{r},t)$ is the phase distribution. The gradient of the phase $S(\mathbf{r},t)$ fixes the velocity of the superfluid through \cite{dalfovo1999theory}
\begin{equation}
\mathbf{v}(\mathbf{r},t) = (\hbar/m) \bm{\nabla} S(\mathbf{r},t). \label{eq:velocityfield}
\end{equation}
From Eq.~(\ref{eq:velocityfield}) it is easy to see that manipulating the phase of a BEC allows to spatially and temporally control its velocity field. Such manipulation can be achieved by exploiting the dipole potential $U(\textbf{r})=-1/(2\varepsilon_0c)\Re(\eta) I(\textbf{r})$, where $\eta$ is the complex polarizability of the atom, $c$ the speed of light, $\varepsilon_0$ the vacuum permittivity and $I(\textbf{r})$ is the intensity distribution of the "imprinting" laser beam, which can be arbitrarily manipulated using e.g. a spatial light modulator \cite{meyer2017observation}. 
This locally sets the phase of the BEC as $S(\textbf{r})=U(\textbf{r})\tau/\hbar$, where $\tau$ is the illumination time. By suitably adjusting the length of the pulse $\tau$ and the intensity of the "imprinting" laser beam $I(\textbf{r})$, it is possible to imprint on the condensate arbitrarily shaped phase pattern.

\begin{figure}[t]
	\includegraphics[width=\textwidth]{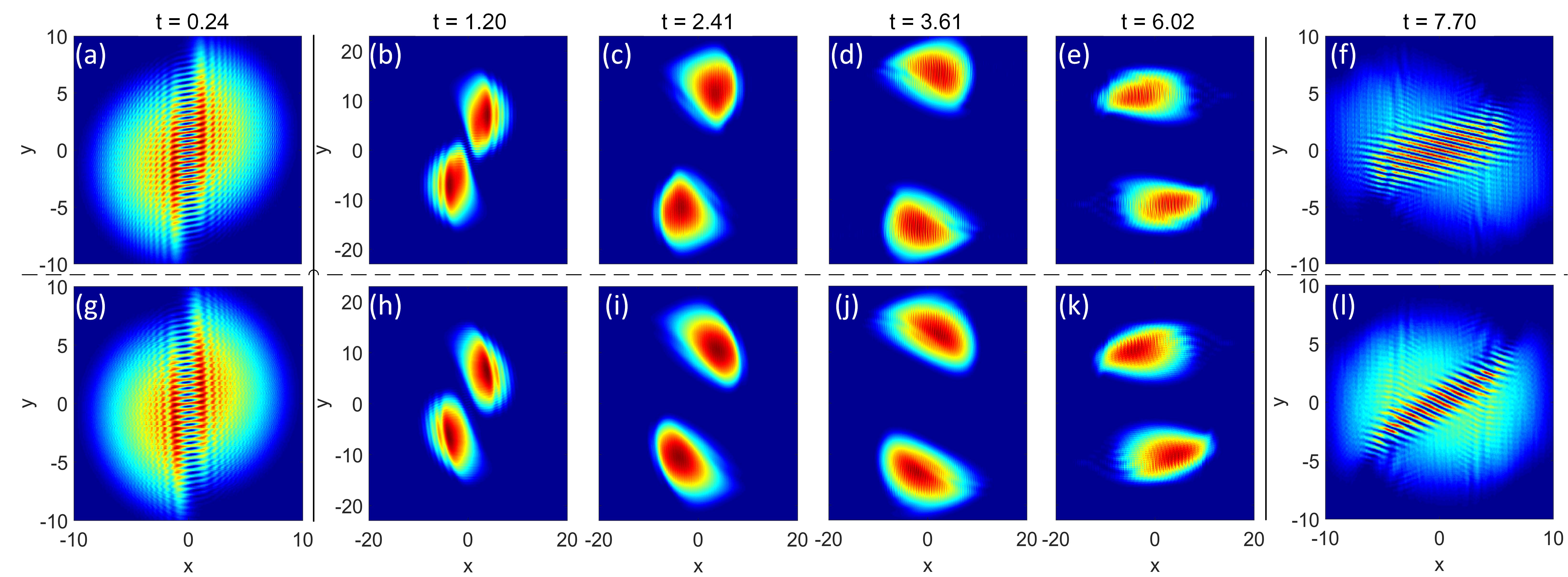}
	\caption{Column density profiles of the BEC at different evolution times $t$ after the phase imprinting. (a)-(f) shows the dynamics after the imprinting of Eq.~(\ref{eq:imprintvelocity}), while (g)-(l) of Eq.~(\ref{eq:imprintedphase}). Note that for t=0.24 and 7.7 the axis range is reduced to better show the initial formation of the vortex conveyor belt and the final interference patterns.}
\label{fig2}
\end{figure}

Let us now consider a BEC trapped in a harmonic potential with frequencies $(\omega_x,\omega_y,\omega_z)$ where $z$ is the direction of propagation of the imprinting beam. To optimize the imprinting, it is convenient to set $\omega_z\gg\omega_x,\omega_y$, leading to a pancake-shaped cloud. Throughout this work, unless otherwise stated, we will express distances in units of the harmonic oscillator length $a_{ho}=\sqrt{\hbar/m\omega_{ho}}$, time in units of $\omega_{ho}^{-1}=(\omega_x\omega_y\omega_z)^{-1/3}$ and energies in units of $\hbar\omega_{ho}$. To illustrate a specific realistic case we consider a BEC of $3\times10^4$ $^{87}$Rb atoms with $\omega_x=\omega_y= 0.06 \omega_z$. We numerically simulate the dynamics of our trapped interferometer starting by imprinting, at $t=0$, two opposite velocities on the two halves of the condensate using the following pattern:
\begin{equation}
\varphi_1(x,y)=\left\{ \begin{array}{ll}
-2\pi y/l & \textrm{if $x < 0$}\\
2\pi y/l & \textrm{if $x \geq 0$}
\end{array} \right.,
\label{eq:imprintvelocity}
\end{equation}
where the coefficient $l$ fixes the phase gradient and therefore the induced velocity field. In Fig.~\ref{fig1}(a) we show the phase pattern resulting from imprinting $\varphi_1$ on the BEC while the corresponding velocity field is reported in Fig.~\ref{fig1}(c). The periodicity of the phase pattern along the $y$-axis is $l$. We simulate the evolution following the phase imprinting by solving Eq.~(\ref{eq:GPE}) using a standard split-step Fourier algorithm. The resulting dynamics for the column density distribution (which is often the observable in BEC experiments) is shown in Fig.~\ref{fig2}. The BEC is first split into two clouds that separate and recombine due to the effect of the harmonic trap. When recombining, after approximately half period, a clear interference pattern is formed in the density profile of the BEC.

We first analyse in detail the dynamics immediately following the initial imprinting. We observe that the initially uniform density distribution rapidly decays into a regular line of vortices pinned at $x=0$ and separated by $l/2$ (first panel of Fig.~\ref{fig2}(a)). This behaviour can be understood by looking at the imprinting pattern in Fig.~\ref{fig1}(a), that presents a periodicity of $l$ along $y$ and a jump across $x=0$. The vortex line then behaves as a conveyor belt for the matter wave (each vortex corresponds to a pulley in a belt conveyor system) launching the two halves of the BEC in opposite directions. To better characterize the splitting and launching action of the conveyor belt and isolate its contribution to the overall dynamics, we have performed a series of numerical simulations where we directly imprint such vortex string on the BEC. A vortex is a topological feature of a superfluid characterized by the fact that, in a closed path around the vortex, the phase of the condensate wave function undergoes a 2$\pi$ winding \cite{matthews1999vortices}, which is due to the single-valuedness of the condensate wave function. Therefore, to directly generate the vortex conveyor belt we imprint on the BEC the following pattern:
\begin{eqnarray}
\varphi_2(x,y) &=& \sum_{k=1}^N 2\Big( \arctan \frac{y + (2k-1)d}{\sqrt{x^2 + \big(y + (2k-1)d \big)^2} + x} + {}\nonumber\\
&&{} + \arctan \frac{y - (2k-1)d}{\sqrt{x^2 + \big(y - (2k-1)d \big)^2} + x} \Big), \label{eq:imprintedphase}
\end{eqnarray}
where $N$ is a positive integer, and $d$ is a positive constant. This phase pattern therefore creates singly quantized vortices along the $x=0$ line with an initial distance of $2d$ between them. Fig.~\ref{fig1}(b) shows an example of the imprinted phase pattern and Fig.~\ref{fig1}(d) shows the corresponding velocity field. Comparing Fig.~\ref{fig1}(c) and Fig.~\ref{fig1}(d), we note that when $d=l/4$ the imprinted velocity fields are essentially equivalent for $|x|$ much larger than the healing length. On the contrary, close to $x=0$ the phase pattern resulting from $\varphi_2$ is clearly distorted and intrinsically presents vorticity, that instead is absent in the phase pattern generated by $\varphi_1$. By comparing the evolution following the two imprinting patterns with the condition $d=l/4$  shown in Fig.~\ref{fig2}, it is possible to appreciate the similarity between the two cases, confirming that the vortex conveyor belt appearing after the imprinting of $\varphi_1$ is responsible for the splitting of the condensate and launching of the two halves in opposite directions. The main difference is represented by the fact that excess vorticity generated by $\varphi_2$ induces an additional small rotation to the two separating clouds that results in a more distorted phase pattern. 

\begin{figure}[t]
\centering
	\includegraphics[width=0.6\textwidth]{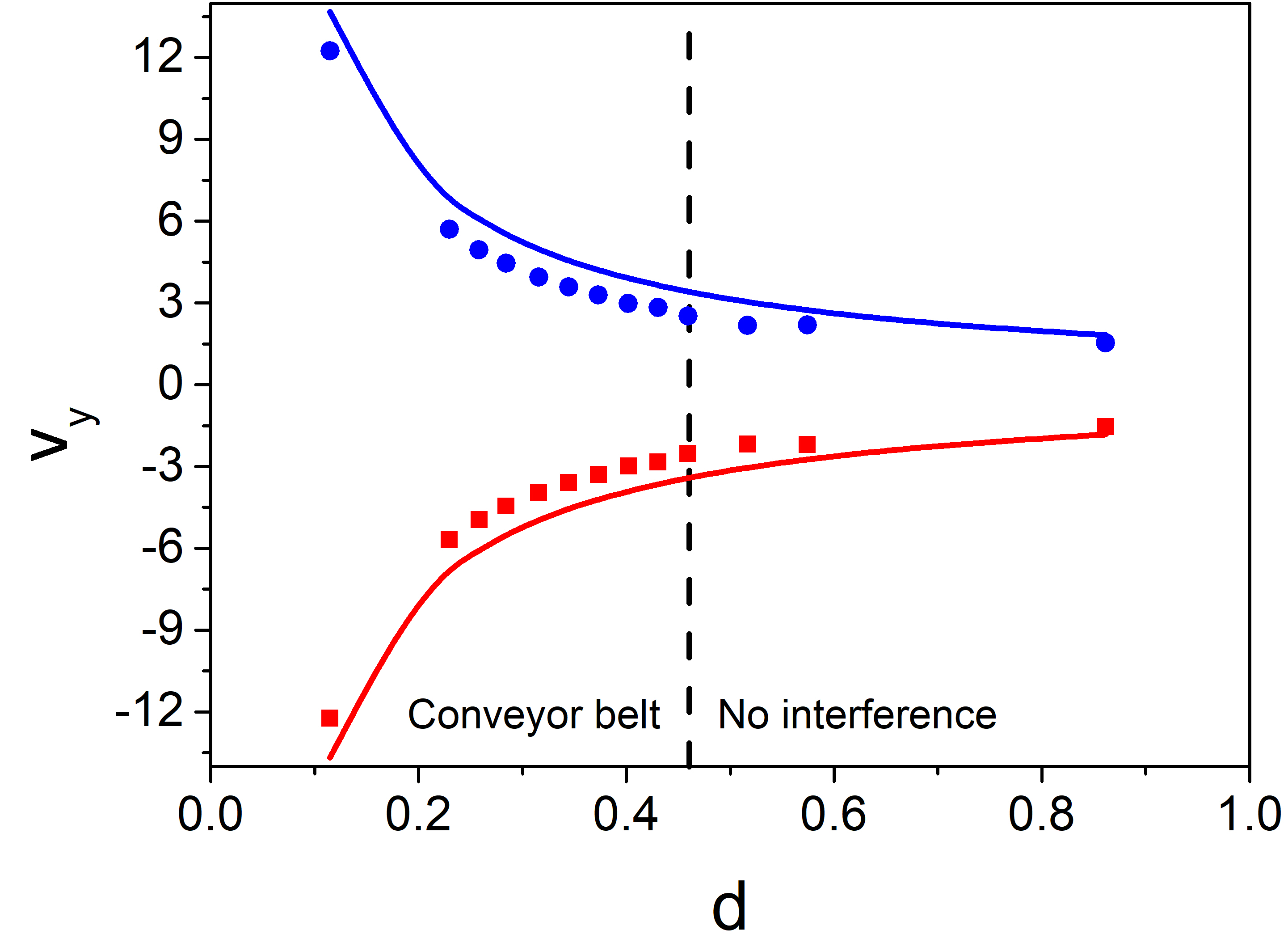}
	\caption{Initial exit velocity along the $y$-axis $v_y$ vs. half initial vortex distance $d$ after the imprinting of the pattern in Eq.~(\ref{eq:imprintedphase}). The blue circles and the red squares represent the values obtained by fitting the simulated density profiles with the procedure explained in the text. The solid lines are calculated with $v_y=d\varphi_1/dy$ with $l=4d$.}
\label{fig3}
\end{figure}

Clearly, due to the similarity of the phase patterns $\varphi_1$ and $\varphi_2$ when $d=l/4$, the initial exit velocity along the $y$-axis is inversely proportional to the distance between the vortices. To study this influence, we performed a series of simulations with different vortex distances. The initial exit velocity along the $y$-axis $v_y$ was extracted from the column density profiles with two steps. First, the centre of mass  positions of the two clouds were determined by fitting the column density profiles with two 2D Gaussian profiles. Due to the existence of the harmonic trap, the centre of mass motion of the clouds is almost harmonic. Therefore, in the second step, we fit the centre of mass motion with a cosine function to obtain the initial exit velocity. Fig.~\ref{fig3} illustrates the result we obtained, confirming that the initial exit velocity along the $y$-axis is inversely proportional to the vortex distance $2d$. For comparison, in Fig.~\ref{fig3} we also plot the curves corresponding to $v_y=d\varphi_1/dy$ with $l=4d$, which correspond to the exit velocities expected imprinting $\varphi_1$. The discrepancy between the velocities obtained with the two methods is again due to the small additional rotation generated by $\varphi_2$. Fig.~\ref{fig3} shows that the energy injected into the BEC decreases with an increasing initial vortex distance. In fact, the energy injected can also be compared by looking at the number and distribution of the vortices ---the energy of a single vortex is higher if it's closer to the centre of the BEC. If we keep increasing the vortex distance, the scheme eventually breaks down as the conveyor belt does not have enough 'pulleys' to efficiently and uniformly split and then launch the two halves of the BEC. We found that for $d \geq 0.46$, it is not possible to generate an interference pattern at the end of the interferometric sequence. In these cases, the vortices created sit stably within the condensate. On the opposite extreme, there is no lower boundary for the vortex distance, meaning that multiply charged vortices can be generated launching the two clouds with increasing exit velocities. 

Let us now characterize the performance of the full interferometric scheme and in particular its ability to detect inertial forces. To this end, we add a constant acceleration $\beta$ towards the negative direction of the $y$-axis. To demonstrate the potential of our scheme for applications requiring large dynamical ranges, we vary $\beta$ in a range spanning 5 orders of magnitude ---from 10$^{-2}$ to 10$^{-7}$. Even in the presence of the external force, the total potential is still harmonic, but its centre is shifted towards lower $y$ values. As a consequence, the initial position of the BEC changes with the magnitude of the acceleration, while the phase imprinting pattern stays constant. 

\begin{figure}[t]
\centering
	\includegraphics[width=0.6\textwidth]{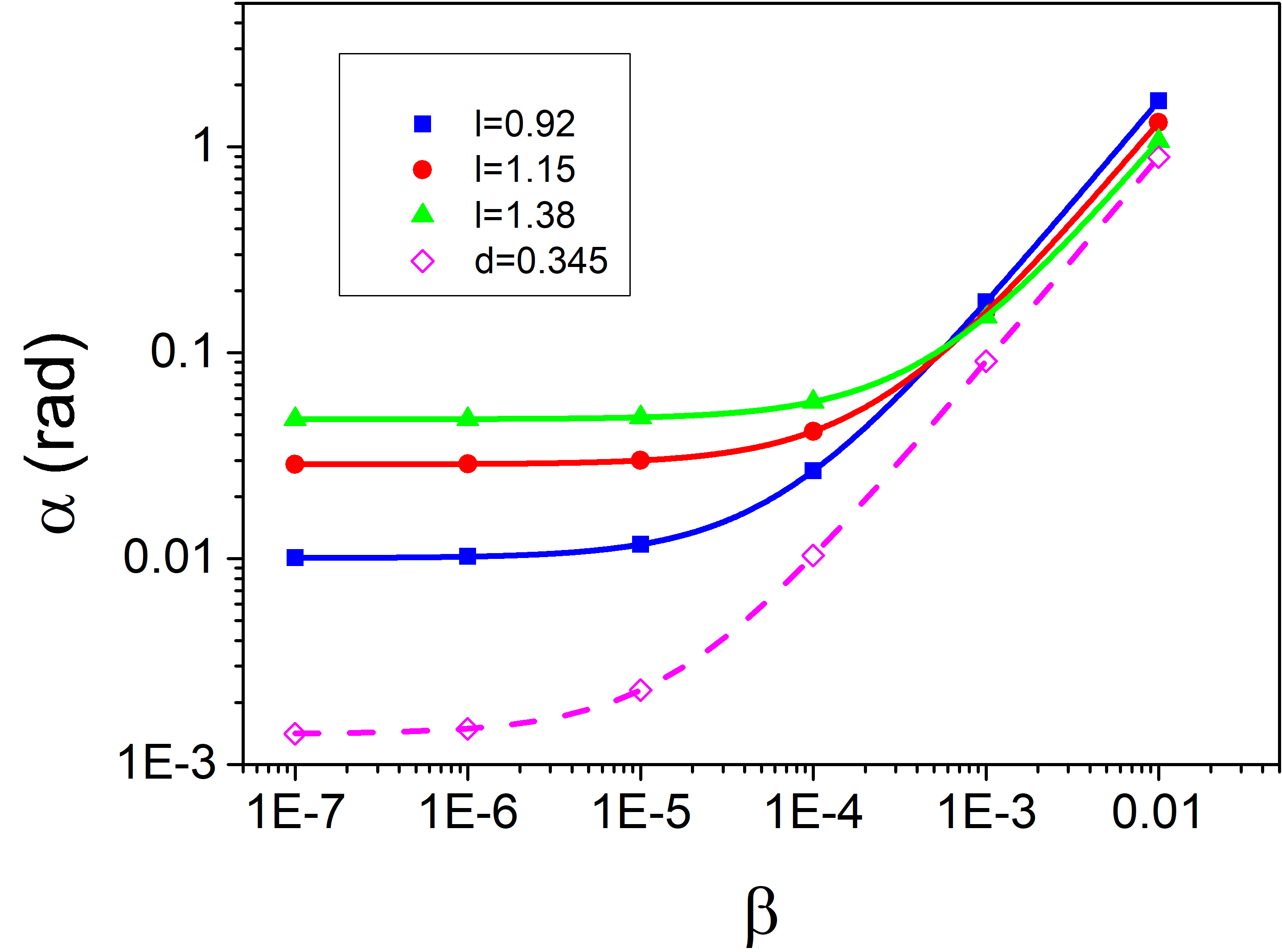}
	\caption{Phase of the interference fringe $\alpha$ vs. acceleration $\beta$ for different values of the phase imprinting coefficient $l$ or $d$. The filled symbols are the results of the numerical simulations after imprinting the pattern in Eq.~(\ref{eq:imprintvelocity}). The solid lines are linear fits to the data. The open points are data extracted from the simulations resulting from the imprint of Eq.~(\ref{eq:imprintedphase}). The dashed line is the corresponding linear fit.}
\label{fig4}
\end{figure}

To measure the acceleration, we fit the density profile of the recombined clouds at the end of the whole interferometer with the product of a Gaussian and a sine function. From this we extract the phase of the interference fringe $\alpha$, which is sensitive to the external force. The values obtained for different imprinted velocities are displayed in Fig.~\ref{fig4}, together with the corresponding linear fits. It can be seen that the relationship between the phase and the acceleration is perfectly linear, demonstrating the validity of our interferometer. As expected, the sensitivity of the accelerometer, given by the slope of the linear fits in Fig.~\ref{fig4}, can be increased by decreasing $l$. For example, for the simulations shown in Fig. 4, the sensitivity increases from 102 to 166 when $l$ is decreased from 1.38 to 0.92. Our simulations again indicate that when we imprint $\varphi_2$, i.e. the sole vortex conveyor belt, at least for high accelerations the performance is comparable with the one obtained by imprinting $\varphi_1$ with $l=4d$. However, as shown in Fig.~\ref{fig4} the sensitivity is reduced by roughly 10\% (for the curve in figure we obtain 89). To give a specific example, if we choose $(\omega_x,\omega_y,\omega_z)=2\pi\times(15,15,250)$ Hz, which are frequencies that can be easily obtained in standard experiments, the curves in Fig.~\ref{fig4} range between 1 and 10$^{5}$ $\mu$gal.

\section*{Discussion}

We have shown that a trapped BEC can be coherently split with a simple phase pattern that generates a vortex conveyor belt. The conveyor belt further launches the two halves of the condensate in opposite directions initiating an interferometric sequence driven by the external trapping potential. We have characterized the initial exit velocity of the two wave packets as a function of the distance between the vortices in the conveyor belt, finding the limit for which the scheme breaks down. We have demonstrated that the interferometer generated is sensitive to external accelerations and that the sensitivity can be improved by increasing the corresponding phase gradient of the imprinted phase pattern. Due to the wide dynamical range and the possibility of being implemented in a compact setup, our interferometer has the potential to be developed into compact and high precision devices. Towards its practical implementation, there are many factors that can affect the acceleration sensitivity. For example, the column density profiles of the BEC are often detected optically. In this case, the optical resolution of the imaging system will limit the resolution of the column density profile and therefore the phase extracted from the interference pattern. Additionally, the optical resolution of the phase imprinting system would set a lower boundary for the $l$ we can realize. Besides, any factors that can influence the evolution of the BEC, such as the fluctuations of the external potential, may also affect the acceleration sensitivity. 

\section*{Methods}

\textbf{Numerical simulations.} For every value of the external acceleration, we first find the ground state of the condensate by finding the solutions of the corresponding time-independent Gross-Pitaevskii equation. This is achieved by iterating in the imaginary time along the steepest descent of the energy. Once the ground state is obtained, we use a standard split-step Fourier algorithm to simulate the dynamics governed by Eq.~(\ref{eq:GPE}). In our simulations, we employ a grid of $(512\times 512\times 8)$ points, except when we consider the imprinting of Eq.~(\ref{eq:imprintvelocity}) with $l \leq 1.15$ or the imprinting of Eq.~(\ref{eq:imprintedphase}) with $d<0.28$, in which cases a grid of $(512\times 1024\times 8)$ points is employed to ensure that the atoms are always in the calculation area while the spatial resolution stays the same.

\noindent \textbf{Interference fringes.} We fit the column density profiles at the end of the interferometric sequence with the following equation:
\begin{equation}
D = A\exp \big( -a_1(x-x_0)^2- 2b_1(x-x_0)(y-y_0)-c_1(y-y_0)^2 \big) \cdot \Big( 1+ \cos \big( a_2(x \cos \theta_1 +y \sin \theta_1) + \alpha \big) \Big) +B, \label{eq:interference}
\end{equation}
where $A$ describes the amplitute and $B$ is the offset. It mainly includes a 2D elliptical Gaussian function times a function that describes an ideal interference pattern. In the Gaussian function, $x_0$ and $y_0$ show the center of mass position of the BEC, while $a_1$, $b_1$ and $c_1$ fix the shape and size of the Gaussian distribution. In the function that describes the interference, $a_2$ is the fringe spatial frequency, $\theta_1$ shows the angle of the interference fringe, and $\alpha$ is the phase of the interference fringe. An example of our fitting procedure is shown in the Supplementary Fig. S1.

\section*{Data Availability}

The datasets generated during the current study are available from the corresponding author on reasonable request.

\bibliography{ref}

\begin{thebibliography}{39}
\expandafter\ifx\csname natexlab\endcsname\relax\def\natexlab#1{#1}\fi
\expandafter\ifx\csname bibnamefont\endcsname\relax
  \def\bibnamefont#1{#1}\fi
\expandafter\ifx\csname bibfnamefont\endcsname\relax
  \def\bibfnamefont#1{#1}\fi
\expandafter\ifx\csname citenamefont\endcsname\relax
  \def\citenamefont#1{#1}\fi
\expandafter\ifx\csname url\endcsname\relax
  \def\url#1{\texttt{#1}}\fi
\expandafter\ifx\csname urlprefix\endcsname\relax\def\urlprefix{URL }\fi
\providecommand{\bibinfo}[2]{#2}
\providecommand{\eprint}[2][]{\url{#2}}

\bibitem[{\citenamefont{Parker et~al.}(2018)\citenamefont{Parker, Yu, Zhong,
  Estey, and M{\"u}ller}}]{parker2018measurement}
\bibinfo{author}{\bibfnamefont{R.~H.} \bibnamefont{Parker}},
  \bibinfo{author}{\bibfnamefont{C.}~\bibnamefont{Yu}},
  \bibinfo{author}{\bibfnamefont{W.}~\bibnamefont{Zhong}},
  \bibinfo{author}{\bibfnamefont{B.}~\bibnamefont{Estey}}, \bibnamefont{and}
  \bibinfo{author}{\bibfnamefont{H.}~\bibnamefont{M{\"u}ller}},
  \bibinfo{journal}{Science} \textbf{\bibinfo{volume}{360}},
  \bibinfo{pages}{191} (\bibinfo{year}{2018}).

\bibitem[{\citenamefont{Thomas et~al.}(2017)\citenamefont{Thomas, Ziane, Pinot,
  Karcher, Imanaliev, Dos~Santos, Merlet, Piquemal, and
  Espel}}]{thomas2017determination}
\bibinfo{author}{\bibfnamefont{M.}~\bibnamefont{Thomas}},
  \bibinfo{author}{\bibfnamefont{D.}~\bibnamefont{Ziane}},
  \bibinfo{author}{\bibfnamefont{P.}~\bibnamefont{Pinot}},
  \bibinfo{author}{\bibfnamefont{R.}~\bibnamefont{Karcher}},
  \bibinfo{author}{\bibfnamefont{A.}~\bibnamefont{Imanaliev}},
  \bibinfo{author}{\bibfnamefont{F.~P.} \bibnamefont{Dos~Santos}},
  \bibinfo{author}{\bibfnamefont{S.}~\bibnamefont{Merlet}},
  \bibinfo{author}{\bibfnamefont{F.}~\bibnamefont{Piquemal}}, \bibnamefont{and}
  \bibinfo{author}{\bibfnamefont{P.}~\bibnamefont{Espel}},
  \bibinfo{journal}{Metrologia} \textbf{\bibinfo{volume}{54}},
  \bibinfo{pages}{468} (\bibinfo{year}{2017}).

\bibitem[{\citenamefont{Dimopoulos et~al.}(2007)\citenamefont{Dimopoulos,
  Graham, Hogan, and Kasevich}}]{dimopoulos2007testing}
\bibinfo{author}{\bibfnamefont{S.}~\bibnamefont{Dimopoulos}},
  \bibinfo{author}{\bibfnamefont{P.~W.} \bibnamefont{Graham}},
  \bibinfo{author}{\bibfnamefont{J.~M.} \bibnamefont{Hogan}}, \bibnamefont{and}
  \bibinfo{author}{\bibfnamefont{M.~A.} \bibnamefont{Kasevich}},
  \bibinfo{journal}{Physical Review Letters} \textbf{\bibinfo{volume}{98}},
  \bibinfo{pages}{111102} (\bibinfo{year}{2007}).

\bibitem[{\citenamefont{Rosi et~al.}(2017)\citenamefont{Rosi, DAmico,
  Cacciapuoti, Sorrentino, Prevedelli, Zych, Brukner, and
  Tino}}]{rosi2017quantum}
\bibinfo{author}{\bibfnamefont{G.}~\bibnamefont{Rosi}},
  \bibinfo{author}{\bibfnamefont{G.}~\bibnamefont{DAmico}},
  \bibinfo{author}{\bibfnamefont{L.}~\bibnamefont{Cacciapuoti}},
  \bibinfo{author}{\bibfnamefont{F.}~\bibnamefont{Sorrentino}},
  \bibinfo{author}{\bibfnamefont{M.}~\bibnamefont{Prevedelli}},
  \bibinfo{author}{\bibfnamefont{M.}~\bibnamefont{Zych}},
  \bibinfo{author}{\bibfnamefont{{\v{C}}.}~\bibnamefont{Brukner}},
  \bibnamefont{and} \bibinfo{author}{\bibfnamefont{G.}~\bibnamefont{Tino}},
  \bibinfo{journal}{Nature communications} \textbf{\bibinfo{volume}{8}},
  \bibinfo{pages}{15529} (\bibinfo{year}{2017}).

\bibitem[{\citenamefont{Dimopoulos et~al.}(2008)\citenamefont{Dimopoulos,
  Graham, Hogan, Kasevich, and Rajendran}}]{dimopoulos2008atomic}
\bibinfo{author}{\bibfnamefont{S.}~\bibnamefont{Dimopoulos}},
  \bibinfo{author}{\bibfnamefont{P.~W.} \bibnamefont{Graham}},
  \bibinfo{author}{\bibfnamefont{J.~M.} \bibnamefont{Hogan}},
  \bibinfo{author}{\bibfnamefont{M.~A.} \bibnamefont{Kasevich}},
  \bibnamefont{and}
  \bibinfo{author}{\bibfnamefont{S.}~\bibnamefont{Rajendran}},
  \bibinfo{journal}{Physical Review D} \textbf{\bibinfo{volume}{78}},
  \bibinfo{pages}{122002} (\bibinfo{year}{2008}).

\bibitem[{\citenamefont{Kovachy et~al.}(2015)\citenamefont{Kovachy, Asenbaum,
  Overstreet, Donnelly, Dickerson, Sugarbaker, Hogan, and
  Kasevich}}]{kovachy2015quantum}
\bibinfo{author}{\bibfnamefont{T.}~\bibnamefont{Kovachy}},
  \bibinfo{author}{\bibfnamefont{P.}~\bibnamefont{Asenbaum}},
  \bibinfo{author}{\bibfnamefont{C.}~\bibnamefont{Overstreet}},
  \bibinfo{author}{\bibfnamefont{C.}~\bibnamefont{Donnelly}},
  \bibinfo{author}{\bibfnamefont{S.}~\bibnamefont{Dickerson}},
  \bibinfo{author}{\bibfnamefont{A.}~\bibnamefont{Sugarbaker}},
  \bibinfo{author}{\bibfnamefont{J.}~\bibnamefont{Hogan}}, \bibnamefont{and}
  \bibinfo{author}{\bibfnamefont{M.}~\bibnamefont{Kasevich}},
  \bibinfo{journal}{Nature} \textbf{\bibinfo{volume}{528}},
  \bibinfo{pages}{530} (\bibinfo{year}{2015}).

\bibitem[{\citenamefont{Canuel et~al.}(2018)\citenamefont{Canuel, Bertoldi,
  Amand, Di~Borgo, Chantrait, Danquigny, {\'A}lvarez, Fang, Freise, Geiger
  et~al.}}]{canuel2018exploring}
\bibinfo{author}{\bibfnamefont{B.}~\bibnamefont{Canuel}},
  \bibinfo{author}{\bibfnamefont{A.}~\bibnamefont{Bertoldi}},
  \bibinfo{author}{\bibfnamefont{L.}~\bibnamefont{Amand}},
  \bibinfo{author}{\bibfnamefont{E.~P.} \bibnamefont{Di~Borgo}},
  \bibinfo{author}{\bibfnamefont{T.}~\bibnamefont{Chantrait}},
  \bibinfo{author}{\bibfnamefont{C.}~\bibnamefont{Danquigny}},
  \bibinfo{author}{\bibfnamefont{M.~D.} \bibnamefont{{\'A}lvarez}},
  \bibinfo{author}{\bibfnamefont{B.}~\bibnamefont{Fang}},
  \bibinfo{author}{\bibfnamefont{A.}~\bibnamefont{Freise}},
  \bibinfo{author}{\bibfnamefont{R.}~\bibnamefont{Geiger}},
  \bibnamefont{et~al.}, \bibinfo{journal}{Scientific Reports}
  \textbf{\bibinfo{volume}{8}}, \bibinfo{pages}{14064} (\bibinfo{year}{2018}).

\bibitem[{\citenamefont{M{\'e}noret et~al.}(2017)\citenamefont{M{\'e}noret,
  Geiger, Stern, Cheinet, Battelier, Zahzam, Dos~Santos, Bresson, Landragin,
  and Bouyer}}]{menoret2017transportable}
\bibinfo{author}{\bibfnamefont{V.}~\bibnamefont{M{\'e}noret}},
  \bibinfo{author}{\bibfnamefont{R.}~\bibnamefont{Geiger}},
  \bibinfo{author}{\bibfnamefont{G.}~\bibnamefont{Stern}},
  \bibinfo{author}{\bibfnamefont{P.}~\bibnamefont{Cheinet}},
  \bibinfo{author}{\bibfnamefont{B.}~\bibnamefont{Battelier}},
  \bibinfo{author}{\bibfnamefont{N.}~\bibnamefont{Zahzam}},
  \bibinfo{author}{\bibfnamefont{F.~P.} \bibnamefont{Dos~Santos}},
  \bibinfo{author}{\bibfnamefont{A.}~\bibnamefont{Bresson}},
  \bibinfo{author}{\bibfnamefont{A.}~\bibnamefont{Landragin}},
  \bibnamefont{and} \bibinfo{author}{\bibfnamefont{P.}~\bibnamefont{Bouyer}},
  in \emph{\bibinfo{booktitle}{International Conference on Space Optics—ICSO
  2010}} (\bibinfo{organization}{International Society for Optics and
  Photonics}, \bibinfo{year}{2017}), vol. \bibinfo{volume}{10565}, p.
  \bibinfo{pages}{1056530}.

\bibitem[{\citenamefont{M{\'e}noret et~al.}(2018)\citenamefont{M{\'e}noret,
  Vermeulen, Le~Moigne, Bonvalot, Bouyer, Landragin, and
  Desruelle}}]{menoret2018gravity}
\bibinfo{author}{\bibfnamefont{V.}~\bibnamefont{M{\'e}noret}},
  \bibinfo{author}{\bibfnamefont{P.}~\bibnamefont{Vermeulen}},
  \bibinfo{author}{\bibfnamefont{N.}~\bibnamefont{Le~Moigne}},
  \bibinfo{author}{\bibfnamefont{S.}~\bibnamefont{Bonvalot}},
  \bibinfo{author}{\bibfnamefont{P.}~\bibnamefont{Bouyer}},
  \bibinfo{author}{\bibfnamefont{A.}~\bibnamefont{Landragin}},
  \bibnamefont{and}
  \bibinfo{author}{\bibfnamefont{B.}~\bibnamefont{Desruelle}},
  \bibinfo{journal}{Scientific reports} \textbf{\bibinfo{volume}{8}},
  \bibinfo{pages}{12300} (\bibinfo{year}{2018}).

\bibitem[{\citenamefont{Giltner et~al.}(1995)\citenamefont{Giltner, McGowan,
  and Lee}}]{giltner1995atom}
\bibinfo{author}{\bibfnamefont{D.~M.} \bibnamefont{Giltner}},
  \bibinfo{author}{\bibfnamefont{R.~W.} \bibnamefont{McGowan}},
  \bibnamefont{and} \bibinfo{author}{\bibfnamefont{S.~A.} \bibnamefont{Lee}},
  \bibinfo{journal}{Physical Review Letters} \textbf{\bibinfo{volume}{75}},
  \bibinfo{pages}{2638} (\bibinfo{year}{1995}).

\bibitem[{\citenamefont{Keith et~al.}(1991)\citenamefont{Keith, Ekstrom,
  Turchette, and Pritchard}}]{keith1991interferometer}
\bibinfo{author}{\bibfnamefont{D.~W.} \bibnamefont{Keith}},
  \bibinfo{author}{\bibfnamefont{C.~R.} \bibnamefont{Ekstrom}},
  \bibinfo{author}{\bibfnamefont{Q.~A.} \bibnamefont{Turchette}},
  \bibnamefont{and} \bibinfo{author}{\bibfnamefont{D.~E.}
  \bibnamefont{Pritchard}}, \bibinfo{journal}{Physical Review Letters}
  \textbf{\bibinfo{volume}{66}}, \bibinfo{pages}{2693} (\bibinfo{year}{1991}).

\bibitem[{\citenamefont{Kasevich and Chu}(1991)}]{kasevich1991atomic}
\bibinfo{author}{\bibfnamefont{M.}~\bibnamefont{Kasevich}} \bibnamefont{and}
  \bibinfo{author}{\bibfnamefont{S.}~\bibnamefont{Chu}},
  \bibinfo{journal}{Physical Review Letters} \textbf{\bibinfo{volume}{67}},
  \bibinfo{pages}{181} (\bibinfo{year}{1991}).

\bibitem[{\citenamefont{Rasel et~al.}(1995)\citenamefont{Rasel, Oberthaler,
  Batelaan, Schmiedmayer, and Zeilinger}}]{rasel1995atom}
\bibinfo{author}{\bibfnamefont{E.~M.} \bibnamefont{Rasel}},
  \bibinfo{author}{\bibfnamefont{M.~K.} \bibnamefont{Oberthaler}},
  \bibinfo{author}{\bibfnamefont{H.}~\bibnamefont{Batelaan}},
  \bibinfo{author}{\bibfnamefont{J.}~\bibnamefont{Schmiedmayer}},
  \bibnamefont{and}
  \bibinfo{author}{\bibfnamefont{A.}~\bibnamefont{Zeilinger}},
  \bibinfo{journal}{Physical Review Letters} \textbf{\bibinfo{volume}{75}},
  \bibinfo{pages}{2633} (\bibinfo{year}{1995}).

\bibitem[{\citenamefont{Freier et~al.}(2016)\citenamefont{Freier, Hauth,
  Schkolnik, Leykauf, Schilling, Wziontek, Scherneck, M{\"u}ller, and
  Peters}}]{freier2016mobile}
\bibinfo{author}{\bibfnamefont{C.}~\bibnamefont{Freier}},
  \bibinfo{author}{\bibfnamefont{M.}~\bibnamefont{Hauth}},
  \bibinfo{author}{\bibfnamefont{V.}~\bibnamefont{Schkolnik}},
  \bibinfo{author}{\bibfnamefont{B.}~\bibnamefont{Leykauf}},
  \bibinfo{author}{\bibfnamefont{M.}~\bibnamefont{Schilling}},
  \bibinfo{author}{\bibfnamefont{H.}~\bibnamefont{Wziontek}},
  \bibinfo{author}{\bibfnamefont{H.-G.} \bibnamefont{Scherneck}},
  \bibinfo{author}{\bibfnamefont{J.}~\bibnamefont{M{\"u}ller}},
  \bibnamefont{and} \bibinfo{author}{\bibfnamefont{A.}~\bibnamefont{Peters}},
  in \emph{\bibinfo{booktitle}{Journal of Physics: Conference Series}}
  (\bibinfo{organization}{IOP Publishing}, \bibinfo{year}{2016}), vol.
  \bibinfo{volume}{723}, p. \bibinfo{pages}{012050}.

\bibitem[{\citenamefont{Dickerson et~al.}(2013)\citenamefont{Dickerson, Hogan,
  Sugarbaker, Johnson, and Kasevich}}]{dickerson2013multiaxis}
\bibinfo{author}{\bibfnamefont{S.~M.} \bibnamefont{Dickerson}},
  \bibinfo{author}{\bibfnamefont{J.~M.} \bibnamefont{Hogan}},
  \bibinfo{author}{\bibfnamefont{A.}~\bibnamefont{Sugarbaker}},
  \bibinfo{author}{\bibfnamefont{D.~M.} \bibnamefont{Johnson}},
  \bibnamefont{and} \bibinfo{author}{\bibfnamefont{M.~A.}
  \bibnamefont{Kasevich}}, \bibinfo{journal}{Physical Review Letters}
  \textbf{\bibinfo{volume}{111}}, \bibinfo{pages}{083001}
  (\bibinfo{year}{2013}).

\bibitem[{\citenamefont{Hu et~al.}(2013)\citenamefont{Hu, Sun, Duan, Zhou,
  Chen, Zhan, Zhang, and Luo}}]{hu2013demonstration}
\bibinfo{author}{\bibfnamefont{Z.-K.} \bibnamefont{Hu}},
  \bibinfo{author}{\bibfnamefont{B.-L.} \bibnamefont{Sun}},
  \bibinfo{author}{\bibfnamefont{X.-C.} \bibnamefont{Duan}},
  \bibinfo{author}{\bibfnamefont{M.-K.} \bibnamefont{Zhou}},
  \bibinfo{author}{\bibfnamefont{L.-L.} \bibnamefont{Chen}},
  \bibinfo{author}{\bibfnamefont{S.}~\bibnamefont{Zhan}},
  \bibinfo{author}{\bibfnamefont{Q.-Z.} \bibnamefont{Zhang}}, \bibnamefont{and}
  \bibinfo{author}{\bibfnamefont{J.}~\bibnamefont{Luo}},
  \bibinfo{journal}{Physical Review A} \textbf{\bibinfo{volume}{88}},
  \bibinfo{pages}{043610} (\bibinfo{year}{2013}).

\bibitem[{\citenamefont{Gauguet et~al.}(2009)\citenamefont{Gauguet, Canuel,
  L{\'e}v{\`e}que, Chaibi, and Landragin}}]{gauguet2009characterization}
\bibinfo{author}{\bibfnamefont{A.}~\bibnamefont{Gauguet}},
  \bibinfo{author}{\bibfnamefont{B.}~\bibnamefont{Canuel}},
  \bibinfo{author}{\bibfnamefont{T.}~\bibnamefont{L{\'e}v{\`e}que}},
  \bibinfo{author}{\bibfnamefont{W.}~\bibnamefont{Chaibi}}, \bibnamefont{and}
  \bibinfo{author}{\bibfnamefont{A.}~\bibnamefont{Landragin}},
  \bibinfo{journal}{Physical Review A} \textbf{\bibinfo{volume}{80}},
  \bibinfo{pages}{063604} (\bibinfo{year}{2009}).

\bibitem[{\citenamefont{Wu et~al.}(2007)\citenamefont{Wu, Su, and
  Prentiss}}]{wu2007demonstration}
\bibinfo{author}{\bibfnamefont{S.}~\bibnamefont{Wu}},
  \bibinfo{author}{\bibfnamefont{E.}~\bibnamefont{Su}}, \bibnamefont{and}
  \bibinfo{author}{\bibfnamefont{M.}~\bibnamefont{Prentiss}},
  \bibinfo{journal}{Physical Review Letters} \textbf{\bibinfo{volume}{99}},
  \bibinfo{pages}{173201} (\bibinfo{year}{2007}).

\bibitem[{\citenamefont{Schumm et~al.}(2005)\citenamefont{Schumm, Hofferberth,
  Andersson, Wildermuth, Groth, Bar-Joseph, Schmiedmayer, and
  Kr{\"u}ger}}]{schumm2005matter}
\bibinfo{author}{\bibfnamefont{T.}~\bibnamefont{Schumm}},
  \bibinfo{author}{\bibfnamefont{S.}~\bibnamefont{Hofferberth}},
  \bibinfo{author}{\bibfnamefont{L.~M.} \bibnamefont{Andersson}},
  \bibinfo{author}{\bibfnamefont{S.}~\bibnamefont{Wildermuth}},
  \bibinfo{author}{\bibfnamefont{S.}~\bibnamefont{Groth}},
  \bibinfo{author}{\bibfnamefont{I.}~\bibnamefont{Bar-Joseph}},
  \bibinfo{author}{\bibfnamefont{J.}~\bibnamefont{Schmiedmayer}},
  \bibnamefont{and}
  \bibinfo{author}{\bibfnamefont{P.}~\bibnamefont{Kr{\"u}ger}},
  \bibinfo{journal}{Nature Physics} \textbf{\bibinfo{volume}{1}},
  \bibinfo{pages}{57} (\bibinfo{year}{2005}).

\bibitem[{\citenamefont{Kasevich}(2002)}]{kasevich2002coherence}
\bibinfo{author}{\bibfnamefont{M.~A.} \bibnamefont{Kasevich}},
  \bibinfo{journal}{Science} \textbf{\bibinfo{volume}{298}},
  \bibinfo{pages}{1363} (\bibinfo{year}{2002}).

\bibitem[{\citenamefont{Kreutzmann et~al.}(2004)\citenamefont{Kreutzmann,
  Poulsen, Lewenstein, Dumke, Ertmer, Birkl, and
  Sanpera}}]{kreutzmann2004coherence}
\bibinfo{author}{\bibfnamefont{H.}~\bibnamefont{Kreutzmann}},
  \bibinfo{author}{\bibfnamefont{U.}~\bibnamefont{Poulsen}},
  \bibinfo{author}{\bibfnamefont{M.}~\bibnamefont{Lewenstein}},
  \bibinfo{author}{\bibfnamefont{R.}~\bibnamefont{Dumke}},
  \bibinfo{author}{\bibfnamefont{W.}~\bibnamefont{Ertmer}},
  \bibinfo{author}{\bibfnamefont{G.}~\bibnamefont{Birkl}}, \bibnamefont{and}
  \bibinfo{author}{\bibfnamefont{A.}~\bibnamefont{Sanpera}},
  \bibinfo{journal}{Physical Review Letters} \textbf{\bibinfo{volume}{92}},
  \bibinfo{pages}{163201} (\bibinfo{year}{2004}).

\bibitem[{\citenamefont{Dimopoulos and Geraci}(2003)}]{dimopoulos2003probing}
\bibinfo{author}{\bibfnamefont{S.}~\bibnamefont{Dimopoulos}} \bibnamefont{and}
  \bibinfo{author}{\bibfnamefont{A.~A.} \bibnamefont{Geraci}},
  \bibinfo{journal}{Physical Review D} \textbf{\bibinfo{volume}{68}},
  \bibinfo{pages}{124021} (\bibinfo{year}{2003}).

\bibitem[{\citenamefont{Alauze et~al.}(2018)\citenamefont{Alauze, Bonnin,
  Solaro, and Dos~Santos}}]{alauze2018trapped}
\bibinfo{author}{\bibfnamefont{X.}~\bibnamefont{Alauze}},
  \bibinfo{author}{\bibfnamefont{A.}~\bibnamefont{Bonnin}},
  \bibinfo{author}{\bibfnamefont{C.}~\bibnamefont{Solaro}}, \bibnamefont{and}
  \bibinfo{author}{\bibfnamefont{F.~P.} \bibnamefont{Dos~Santos}},
  \bibinfo{journal}{New Journal of Physics} \textbf{\bibinfo{volume}{20}},
  \bibinfo{pages}{083014} (\bibinfo{year}{2018}).

\bibitem[{\citenamefont{Burke et~al.}(2008)\citenamefont{Burke, Deissler,
  Hughes, and Sackett}}]{burke2008confinement}
\bibinfo{author}{\bibfnamefont{J.}~\bibnamefont{Burke}},
  \bibinfo{author}{\bibfnamefont{B.}~\bibnamefont{Deissler}},
  \bibinfo{author}{\bibfnamefont{K.}~\bibnamefont{Hughes}}, \bibnamefont{and}
  \bibinfo{author}{\bibfnamefont{C.}~\bibnamefont{Sackett}},
  \bibinfo{journal}{Physical Review A} \textbf{\bibinfo{volume}{78}},
  \bibinfo{pages}{023619} (\bibinfo{year}{2008}).

\bibitem[{\citenamefont{Andrews et~al.}(1997)\citenamefont{Andrews, Townsend,
  Miesner, Durfee, Kurn, and Ketterle}}]{andrews1997observation}
\bibinfo{author}{\bibfnamefont{M.}~\bibnamefont{Andrews}},
  \bibinfo{author}{\bibfnamefont{C.}~\bibnamefont{Townsend}},
  \bibinfo{author}{\bibfnamefont{H.-J.} \bibnamefont{Miesner}},
  \bibinfo{author}{\bibfnamefont{D.}~\bibnamefont{Durfee}},
  \bibinfo{author}{\bibfnamefont{D.}~\bibnamefont{Kurn}}, \bibnamefont{and}
  \bibinfo{author}{\bibfnamefont{W.}~\bibnamefont{Ketterle}},
  \bibinfo{journal}{Science} \textbf{\bibinfo{volume}{275}},
  \bibinfo{pages}{637} (\bibinfo{year}{1997}).

\bibitem[{\citenamefont{Wang et~al.}(2005)\citenamefont{Wang, Anderson, Bright,
  Cornell, Diot, Kishimoto, Prentiss, Saravanan, Segal, and Wu}}]{wang2005atom}
\bibinfo{author}{\bibfnamefont{Y.-J.} \bibnamefont{Wang}},
  \bibinfo{author}{\bibfnamefont{D.~Z.} \bibnamefont{Anderson}},
  \bibinfo{author}{\bibfnamefont{V.~M.} \bibnamefont{Bright}},
  \bibinfo{author}{\bibfnamefont{E.~A.} \bibnamefont{Cornell}},
  \bibinfo{author}{\bibfnamefont{Q.}~\bibnamefont{Diot}},
  \bibinfo{author}{\bibfnamefont{T.}~\bibnamefont{Kishimoto}},
  \bibinfo{author}{\bibfnamefont{M.}~\bibnamefont{Prentiss}},
  \bibinfo{author}{\bibfnamefont{R.}~\bibnamefont{Saravanan}},
  \bibinfo{author}{\bibfnamefont{S.~R.} \bibnamefont{Segal}}, \bibnamefont{and}
  \bibinfo{author}{\bibfnamefont{S.}~\bibnamefont{Wu}},
  \bibinfo{journal}{Physical Review Letters} \textbf{\bibinfo{volume}{94}},
  \bibinfo{pages}{090405} (\bibinfo{year}{2005}).

\bibitem[{\citenamefont{Sapiro et~al.}(2009)\citenamefont{Sapiro, Zhang, and
  Raithel}}]{sapiro2009atom}
\bibinfo{author}{\bibfnamefont{R.}~\bibnamefont{Sapiro}},
  \bibinfo{author}{\bibfnamefont{R.}~\bibnamefont{Zhang}}, \bibnamefont{and}
  \bibinfo{author}{\bibfnamefont{G.}~\bibnamefont{Raithel}},
  \bibinfo{journal}{Physical Review A} \textbf{\bibinfo{volume}{79}},
  \bibinfo{pages}{043630} (\bibinfo{year}{2009}).

\bibitem[{\citenamefont{Shin et~al.}(2004)\citenamefont{Shin, Saba, Pasquini,
  Ketterle, Pritchard, and Leanhardt}}]{shin2004atom}
\bibinfo{author}{\bibfnamefont{Y.}~\bibnamefont{Shin}},
  \bibinfo{author}{\bibfnamefont{M.}~\bibnamefont{Saba}},
  \bibinfo{author}{\bibfnamefont{T.}~\bibnamefont{Pasquini}},
  \bibinfo{author}{\bibfnamefont{W.}~\bibnamefont{Ketterle}},
  \bibinfo{author}{\bibfnamefont{D.}~\bibnamefont{Pritchard}},
  \bibnamefont{and}
  \bibinfo{author}{\bibfnamefont{A.}~\bibnamefont{Leanhardt}},
  \bibinfo{journal}{Physical Review Letters} \textbf{\bibinfo{volume}{92}},
  \bibinfo{pages}{050405} (\bibinfo{year}{2004}).

\bibitem[{\citenamefont{Guarrera et~al.}(2015)\citenamefont{Guarrera, Szmuk,
  Reichel, and Rosenbusch}}]{guarrera}
\bibinfo{author}{\bibfnamefont{V.}~\bibnamefont{Guarrera}},
  \bibinfo{author}{\bibfnamefont{R.}~\bibnamefont{Szmuk}},
  \bibinfo{author}{\bibfnamefont{J.}~\bibnamefont{Reichel}}, \bibnamefont{and}
  \bibinfo{author}{\bibfnamefont{P.}~\bibnamefont{Rosenbusch}},
  \bibinfo{journal}{New Journal of Physics} \textbf{\bibinfo{volume}{17}},
  \bibinfo{pages}{083022} (\bibinfo{year}{2015}),
  \urlprefix\url{http://stacks.iop.org/1367-2630/17/i=8/a=083022}.

\bibitem[{\citenamefont{Riedel et~al.}(2010)\citenamefont{Riedel, B{\"o}hi, Li,
  H{\"a}nsch, Sinatra, and Treutlein}}]{riedel2012atom}
\bibinfo{author}{\bibfnamefont{M.~F.} \bibnamefont{Riedel}},
  \bibinfo{author}{\bibfnamefont{P.}~\bibnamefont{B{\"o}hi}},
  \bibinfo{author}{\bibfnamefont{Y.}~\bibnamefont{Li}},
  \bibinfo{author}{\bibfnamefont{T.~W.} \bibnamefont{H{\"a}nsch}},
  \bibinfo{author}{\bibfnamefont{A.}~\bibnamefont{Sinatra}}, \bibnamefont{and}
  \bibinfo{author}{\bibfnamefont{P.}~\bibnamefont{Treutlein}},
  \bibinfo{journal}{Nature} \textbf{\bibinfo{volume}{464}},
  \bibinfo{pages}{1170} (\bibinfo{year}{2010}).

\bibitem[{\citenamefont{Marchant et~al.}(2013)\citenamefont{Marchant, Billam,
  Wiles, Yu, Gardiner, and Cornish}}]{marchant2013controlled}
\bibinfo{author}{\bibfnamefont{A.}~\bibnamefont{Marchant}},
  \bibinfo{author}{\bibfnamefont{T.}~\bibnamefont{Billam}},
  \bibinfo{author}{\bibfnamefont{T.}~\bibnamefont{Wiles}},
  \bibinfo{author}{\bibfnamefont{M.}~\bibnamefont{Yu}},
  \bibinfo{author}{\bibfnamefont{S.}~\bibnamefont{Gardiner}}, \bibnamefont{and}
  \bibinfo{author}{\bibfnamefont{S.}~\bibnamefont{Cornish}},
  \bibinfo{journal}{Nature Communications} \textbf{\bibinfo{volume}{4}},
  \bibinfo{pages}{1865} (\bibinfo{year}{2013}).

\bibitem[{\citenamefont{McDonald et~al.}(2014)\citenamefont{McDonald, Kuhn,
  Hardman, Bennetts, Everitt, Altin, Debs, Close, and
  Robins}}]{mcdonald2014bright}
\bibinfo{author}{\bibfnamefont{G.~D.} \bibnamefont{McDonald}},
  \bibinfo{author}{\bibfnamefont{C.~C.} \bibnamefont{Kuhn}},
  \bibinfo{author}{\bibfnamefont{K.~S.} \bibnamefont{Hardman}},
  \bibinfo{author}{\bibfnamefont{S.}~\bibnamefont{Bennetts}},
  \bibinfo{author}{\bibfnamefont{P.~J.} \bibnamefont{Everitt}},
  \bibinfo{author}{\bibfnamefont{P.~A.} \bibnamefont{Altin}},
  \bibinfo{author}{\bibfnamefont{J.~E.} \bibnamefont{Debs}},
  \bibinfo{author}{\bibfnamefont{J.~D.} \bibnamefont{Close}}, \bibnamefont{and}
  \bibinfo{author}{\bibfnamefont{N.~P.} \bibnamefont{Robins}},
  \bibinfo{journal}{Physical Review Letters} \textbf{\bibinfo{volume}{113}},
  \bibinfo{pages}{013002} (\bibinfo{year}{2014}).

\bibitem[{\citenamefont{Gajda et~al.}(1999)\citenamefont{Gajda, Lewenstein,
  Sengstock, Birkl, Ertmer et~al.}}]{gajda1999optical}
\bibinfo{author}{\bibfnamefont{M.}~\bibnamefont{Gajda}},
  \bibinfo{author}{\bibfnamefont{M.}~\bibnamefont{Lewenstein}},
  \bibinfo{author}{\bibfnamefont{K.}~\bibnamefont{Sengstock}},
  \bibinfo{author}{\bibfnamefont{G.}~\bibnamefont{Birkl}},
  \bibinfo{author}{\bibfnamefont{W.}~\bibnamefont{Ertmer}},
  \bibnamefont{et~al.}, \bibinfo{journal}{Physical Review A}
  \textbf{\bibinfo{volume}{60}}, \bibinfo{pages}{R3381} (\bibinfo{year}{1999}).

\bibitem[{\citenamefont{Becker et~al.}(2008)\citenamefont{Becker, Stellmer,
  Soltan-Panahi, D{\"o}rscher, Baumert, Richter, Kronj{\"a}ger, Bongs, and
  Sengstock}}]{becker2008oscillations}
\bibinfo{author}{\bibfnamefont{C.}~\bibnamefont{Becker}},
  \bibinfo{author}{\bibfnamefont{S.}~\bibnamefont{Stellmer}},
  \bibinfo{author}{\bibfnamefont{P.}~\bibnamefont{Soltan-Panahi}},
  \bibinfo{author}{\bibfnamefont{S.}~\bibnamefont{D{\"o}rscher}},
  \bibinfo{author}{\bibfnamefont{M.}~\bibnamefont{Baumert}},
  \bibinfo{author}{\bibfnamefont{E.-M.} \bibnamefont{Richter}},
  \bibinfo{author}{\bibfnamefont{J.}~\bibnamefont{Kronj{\"a}ger}},
  \bibinfo{author}{\bibfnamefont{K.}~\bibnamefont{Bongs}}, \bibnamefont{and}
  \bibinfo{author}{\bibfnamefont{K.}~\bibnamefont{Sengstock}},
  \bibinfo{journal}{Nature Physics} \textbf{\bibinfo{volume}{4}},
  \bibinfo{pages}{496} (\bibinfo{year}{2008}).

\bibitem[{\citenamefont{Denschlag et~al.}(2000)\citenamefont{Denschlag,
  Simsarian, Feder, Clark, Collins, Cubizolles, Deng, Hagley, Helmerson,
  Reinhardt et~al.}}]{denschlag2000generating}
\bibinfo{author}{\bibfnamefont{J.}~\bibnamefont{Denschlag}},
  \bibinfo{author}{\bibfnamefont{J.~E.} \bibnamefont{Simsarian}},
  \bibinfo{author}{\bibfnamefont{D.~L.} \bibnamefont{Feder}},
  \bibinfo{author}{\bibfnamefont{C.~W.} \bibnamefont{Clark}},
  \bibinfo{author}{\bibfnamefont{L.~A.} \bibnamefont{Collins}},
  \bibinfo{author}{\bibfnamefont{J.}~\bibnamefont{Cubizolles}},
  \bibinfo{author}{\bibfnamefont{L.}~\bibnamefont{Deng}},
  \bibinfo{author}{\bibfnamefont{E.~W.} \bibnamefont{Hagley}},
  \bibinfo{author}{\bibfnamefont{K.}~\bibnamefont{Helmerson}},
  \bibinfo{author}{\bibfnamefont{W.~P.} \bibnamefont{Reinhardt}},
  \bibnamefont{et~al.}, \bibinfo{journal}{Science}
  \textbf{\bibinfo{volume}{287}}, \bibinfo{pages}{97} (\bibinfo{year}{2000}).

\bibitem[{\citenamefont{Meyer et~al.}(2017)\citenamefont{Meyer, Proud,
  Perea-Ortiz, ONeale, Baumert, Holynski, Kronj{\"a}ger, Barontini, and
  Bongs}}]{meyer2017observation}
\bibinfo{author}{\bibfnamefont{N.}~\bibnamefont{Meyer}},
  \bibinfo{author}{\bibfnamefont{H.}~\bibnamefont{Proud}},
  \bibinfo{author}{\bibfnamefont{M.}~\bibnamefont{Perea-Ortiz}},
  \bibinfo{author}{\bibfnamefont{C.}~\bibnamefont{ONeale}},
  \bibinfo{author}{\bibfnamefont{M.}~\bibnamefont{Baumert}},
  \bibinfo{author}{\bibfnamefont{M.}~\bibnamefont{Holynski}},
  \bibinfo{author}{\bibfnamefont{J.}~\bibnamefont{Kronj{\"a}ger}},
  \bibinfo{author}{\bibfnamefont{G.}~\bibnamefont{Barontini}},
  \bibnamefont{and} \bibinfo{author}{\bibfnamefont{K.}~\bibnamefont{Bongs}},
  \bibinfo{journal}{Physical Review Letters} \textbf{\bibinfo{volume}{119}},
  \bibinfo{pages}{150403} (\bibinfo{year}{2017}).

\bibitem[{\citenamefont{Ruschhaupt et~al.}(2009)\citenamefont{Ruschhaupt,
  Del~Campo, and Muga}}]{ruschhaupt2009momentum}
\bibinfo{author}{\bibfnamefont{A.}~\bibnamefont{Ruschhaupt}},
  \bibinfo{author}{\bibfnamefont{A.}~\bibnamefont{Del~Campo}},
  \bibnamefont{and} \bibinfo{author}{\bibfnamefont{J.}~\bibnamefont{Muga}},
  \bibinfo{journal}{Physical Review A} \textbf{\bibinfo{volume}{79}},
  \bibinfo{pages}{023616} (\bibinfo{year}{2009}).

\bibitem[{\citenamefont{Dalfovo et~al.}(1999)\citenamefont{Dalfovo, Giorgini,
  Pitaevskii, and Stringari}}]{dalfovo1999theory}
\bibinfo{author}{\bibfnamefont{F.}~\bibnamefont{Dalfovo}},
  \bibinfo{author}{\bibfnamefont{S.}~\bibnamefont{Giorgini}},
  \bibinfo{author}{\bibfnamefont{L.~P.} \bibnamefont{Pitaevskii}},
  \bibnamefont{and}
  \bibinfo{author}{\bibfnamefont{S.}~\bibnamefont{Stringari}},
  \bibinfo{journal}{Reviews of Modern Physics} \textbf{\bibinfo{volume}{71}},
  \bibinfo{pages}{463} (\bibinfo{year}{1999}).

\bibitem[{\citenamefont{Matthews et~al.}(1999)\citenamefont{Matthews, Anderson,
  Haljan, Hall, Wieman, and Cornell}}]{matthews1999vortices}
\bibinfo{author}{\bibfnamefont{M.~R.} \bibnamefont{Matthews}},
  \bibinfo{author}{\bibfnamefont{B.~P.} \bibnamefont{Anderson}},
  \bibinfo{author}{\bibfnamefont{P.}~\bibnamefont{Haljan}},
  \bibinfo{author}{\bibfnamefont{D.}~\bibnamefont{Hall}},
  \bibinfo{author}{\bibfnamefont{C.}~\bibnamefont{Wieman}}, \bibnamefont{and}
  \bibinfo{author}{\bibfnamefont{E.~A.} \bibnamefont{Cornell}},
  \bibinfo{journal}{Physical Review Letters} \textbf{\bibinfo{volume}{83}},
  \bibinfo{pages}{2498} (\bibinfo{year}{1999}).

\end{thebibliography}

\section*{Acknowledgements}

J. Liu acknowledges support from the China Scholarship Council.

\section*{Author contributions statement}

J.L. performed the numerical simulations and wrote the manuscript. X.W., J.M.M., A.K. and G.B. participated in the discussion of the results and the writing of the manuscript.

\end{document}